\documentclass{article}

\usepackage{arxiv}

\usepackage[utf8]{inputenc} 
\usepackage[T1]{fontenc}    
\usepackage{hyperref}       
\usepackage{url}            
\usepackage{booktabs}       
\usepackage{amsfonts}       
\usepackage{nicefrac}       
\usepackage{microtype}      
\usepackage{lipsum}
\usepackage{graphicx}
\usepackage{lineno,hyperref}
\usepackage{bm}
\usepackage{caption}
\usepackage{multirow}

\usepackage{framed,multirow}
\usepackage{amssymb}
\usepackage{amsmath}
\usepackage{latexsym}
\usepackage{subcaption}
\usepackage[table,xcdraw]{xcolor}

\usepackage[sort&compress,numbers]{natbib}
\emergencystretch=\maxdimen
\title{Clustering and classification of low-dimensional data in explicit feature map domain: intraoperative pixel-wise diagnosis of adenocarcinoma of a colon in a liver}

\author{Ivica Kopriv$\mathrm{a}^*$}

\author{
 Dario Sitnik \\
  Laboratory for Machine Learning and Knowledge Representation \\ Division of Electronics \\ Ruđer Bošković Institute\\ Bijenička cesta 54, 10000, Zagreb, Croatia\\
  \texttt{dsitnik@irb.hr, dario.sitnik@gmail.com}
   \And
 Ivica Kopriva \\
  Laboratory for Machine Learning and Knowledge Representation \\ Division of Electronics \\ Ruđer Bošković Institute\\ Bijenička cesta 54, 10000, Zagreb, Croatia\\
  \texttt{ikopriva@irb.hr, ikopriva@gmail.com}
  }

\begin{document}
\maketitle
\begin{abstract}
Application of artificial intelligence in medicine brings in highly accurate predictions achieved by complex models, the reasoning of which is hard to interpret. Their generalization ability can be reduced because of the lack of pixel wise annotated images that occurs in frozen section tissue analysis. To partially overcome this gap, this paper explores the approximate explicit feature map (aEFM) transform of low-dimensional data into a low-dimensional subspace in Hilbert space. There, with a modest increase in computational complexity, linear algorithms yield improved performance and keep interpretability. They remain amenable to incremental learning that is not a trivial issue for some nonlinear algorithms. We demonstrate proposed methodology on a very large-scale problem related to intraoperative pixel-wise semantic segmentation and clustering of adenocarcinoma of a colon in a liver. Compared to the results in the input space, logistic classifier achieved statistically significant performance improvements in micro balanced accuracy and F1 score in the amounts of 12.04\% and 12.58\%, respectively. Support vector machine classifier yielded the increase of 8.04\% and 9.41\%. For clustering, increases of 0.79\% and 0.85\% are obtained with ultra large-scale spectral clustering algorithm. Results are supported by a discussion of interpretability using Shapely additive explanation values for predictions of linear classifier in input space and aEFM induced space.
\end{abstract}

\keywords{low-dimensional data\and explicit feature maps\and (non)linear classifiers\and semantic segmentation\and interpretability\and intraoperative diagnosis}

\section{Introduction}
\label{intro}
Artificial intelligence instantiated in terms of deep learning (DL) and neural networks (NN) has achieved a stage of highly accurate diagnoses in various medical imaging modalities, sometimes even comparable to human experts \citep{litjens2016deep, de2018clinically, xing2018deep, hu2018deep,mckinney2020international,rajpurkar2018deep}. That is enabled by advances in computing resources, architectures and expert labeled datasets necessary to train DL/NN algorithms \citep{shad2021designing}. However, high diagnostic performance is achieved by highly complex NN models whose decision-making process is hard to interpret and explain \citep{rudin2019stop,tjoa2021asurvey}. That creates a tension between the model’s performance and interpretability \citep{lundberg2017unified}. When decisions are at high-stakes, it is necessary to explain predictions made by the algorithm. Therefore, creating models that are interpretable in the first place should be preferred over explaining the black box models \citep{rudin2019stop}. In medical task such as intraoperative pixel-wise diagnosis of cancer, there is a lack of publicly available datasets with pixel-wise annotated histopathological images of frozen sections \citep{komura2018machine,sitnik2021cocahis}. Demanding sample collection during surgeries and highly time-consuming effort for expert’s labeling are the two main reasons for lack of annotated data. That can create a generalization problem for complex models, such as convolutional NNs \citep{veta2014breast,litjens2016deep,xing2018deep,qiu2018deep}.

To address outlined challenges, we propose a new approach for low-dimensional data. We apply existing segmentation and linear classification algorithms in a space induced by approximate explicit feature maps (aEFMs). The approach is motivated by the fact that  linear classification algorithms remain interpretable when applied to interpretable features. Thus, if features in transformed space are known analytical functions of the original features, the linear classification algorithm in mapped space should remain interpretable. The approach is also justified theoretically because nonlinear classification algorithms such as support vector machines (SVMs) \citep{huang2006kernel} and logistic regression \citep{hastie2009elements} can be seen as linear counterparts operating in an appropriate feature space \citep{vedaldi2012efficient}. The main reason behind why the computation in feature space $\mathcal{H}$ (a.k.a. Hilbert space) is rarely used in computation is that the feature space is usually infinite dimensional (see Appendix A for the EFM map associated with the Gaussian kernel). We use aEFMs of order $m$, $\phi_m(\bm{x})$, to approximate $\mathcal{H}$ and project data $\bm{x}\in \mathbb{R}^d$ from input space to a subspace $\mathcal{H}_D\subset\mathcal{H}$ where linear classification algorithms of small computational complexity are trained. If $d$ is small, dimensionality of $\mathcal{H}_D$ will not be too large. As an example of the aEFMs associated with the Gaussian and polynomial kernels, it applies (see Appendix A for details):

\begin{equation}
\label{eq:Choose}
D = \binom{d+m}{d}\ .
\end{equation}

\noindent Thus, linear classification in aEFM-induced space will be computationally feasible. Owning to the fact that aEFMs are available in an analytical form, the new features $\phi_m(\bm{x})$ are analytical functions of the original features $\bm{x}$. Hence, if original features are interpretable, the features in mapping induced space should be interpretable as well. We validate the proposed approach on image segmentation of hematoxylin-eosin ($H\&E$) stained frozen sections of adenocarcinoma of a colon in a liver \citep{sitnik2021cocahis,cocahis2021}. The CoCaHis dataset comprises 82 images (approximately 1.44 million of pixels per image) of $H\&E$ stained frozen sections collected intraoperatively from 19 patients. The training set is made of 58 images belonging to 13 patients, and the test set comprises 24 images belonging to 6 patients.  As follows, the training set has approximately 75 million of pixels and classification problem is of the very large scale. Therefore, classifier training must be done in the incremental mode.

In considered application, pixel features in the input data space are red, green, and blue color components, i.e. $d=3$. We calculated Shapely additive values \citep{lundberg2017unified,aas2021explaining} to explain logistic classifier’s predictions for cancerous and non-cancerous pixels in input space and aEFM induced space. Because of the low-dimensionality of the original data and linear character of the classifiers in the aEFM-induced space, we could run the algorithms in the incremental learning mode. That is a nontrivial issue for kernel-based nonlinear algorithms. To emphasize a potentially broader impact of the aEFM-based concept, we also applied ultra-scalable spectral clustering (U-SPEC) algorithm \citep{huang2020ultra} for clustering in the input data space and aEFM-induced space.

We want to point out that nonlinear dimensionality expansion transforms were already used with: independent component analysis in magnetic resonance imaging \citep{ouyang2008band} and multispectral imaging \citep{kopriva2009unsupervised}, nonnegative matrix factorization in positron emission tomography imaging \citep{kopriva2017single-channel} and histopathological image analysis \citep{kopriva2021approx}. Those nonlinear transforms were merely constructed on an \textit{ad hoc} basis. To the best of our knowledge, it is for the first time that, on a strong theoretical basis, nonlinear classification is defined as a linear classification in the feature space and applied in the incremental learning mode to a very large-scale classification problem.

To possibly boost classification performance, we also trained an ensemble of SVM  and logistic classifiers in the input-space and aEFM-induced space. Again, ensemble of classifiers in aEFM-induced feature space achieved statistically significant improvements in mentioned metrics relative to the corresponding performances in the input space. Performances of the ensemble of logistic classifiers are improved in the amounts of 1.05\%, 0.96\% and 0.04\%. For the ensemble of SVM classifiers, performance increases are:  1.47\%, 1.89\% and 1.75\%. Thus, in case of the ensemble of classifiers, performances are improved in smaller amount. Already achieved significant performance of the ensembles in the input space is a reason behind such modest improvement. It is interesting that best accuracy is achieved by a single SVM classifier (77.11\%) or a logistic classifier (77.17\%) in a space induced by the Gaussian aEFM of order $m=2$. Hence, that is practically important because the ensemble training is more computationally complex than a single classifier training. In a summary, formulation of the clustering and classification problems in aEFM-induced space yields statistically significant improvement in performance while predictions remain explainable. We support this statement by discussion of Shapely additive explanation values estimated for individual predictions obtained by linear classifier in input space and aEFM induced space.

\section{Methods and materials}
\label{sec:methods}
\begin{figure}[!ht]
\centering
\includegraphics[width=0.4\textwidth]{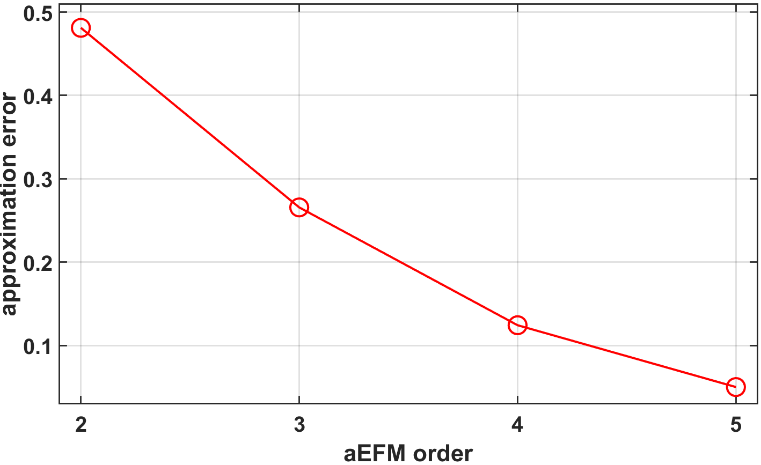}
\caption{ Absolute value of the error between the kernel function $\kappa(\bm{x},\bm{y})=\mathrm{exp} \left (-\left|| \bm{x}-\bm{y}\right||^2 / 2\sigma^2 \right)$ and its inner product-based approximation $\left<\phi(\bm{x}),\phi(\bm{y})\right>_{\mathcal{H}_D}$ as a function of order $m\in\{2,3,4,5\}.$ $\bm{x},\bm{y}\in\mathbb{R}^3_{0+}$ represent two pixels from the color image of $H\&E$ stained frozen section. Pixel values were rescaled to the range [0,1]. Parameter $\sigma$ was set to 0.7071.}
\label{figure:approxErr} 
\end{figure}

For any positive definite kernel function $\kappa(\bm{x},\bm{y})$ there exist a function $\phi(\bm{x})$ that maps the data $\bm{x}$ to a Hilbert space $\mathcal{H}$ such that the kernel trick applies: $\kappa(\bm{x},\bm{y})=\left<\phi(\bm{x}),\phi(\bm{y})\right>_\mathcal{H}$ \citep{scholkopf2002learning}. $\mathcal{H}$ is also known as a feature space, and $\phi(\bm{x})$ as an explicit feature map. In kernel trick-based nonlinear algorithms, the feature maps are not used explicitly. Since feature maps can be infinite dimensional (see Eq.(\ref{eq:appx11}) in Appendix A) for EFM associated with the Gaussian kernel, they are not used too often in computations. However, the approximate EFM of order $m$, $\phi_m(\bm{x})$ , yields good and computationally feasible approximation of kernel function. Hence, the error $\epsilon=\kappa(\bm{x},\bm{y})-\left<\phi(\bm{x}),\phi(\bm{y})\right>_{\mathcal{H}_D}$ is small in magnitude, where $\mathcal{H}_D\subset\mathcal{H}$ is low dimensional subspace induced by $\phi_m(\bm{x})$. The case is illustrated in Fig. \ref{figure:approxErr} for the aEFM associated with the Gaussian kernel. In case of EFM associated with the polynomial kernel, there is no approximation error, i.e. $m$ corresponds to the order of the polynomial.

\subsection{Linear classification and explicit feature maps}
Let us define a dataset as a collection of $N$ data points in $d$-dimensional space, i.e. $\bm{X}:=\left\{\bm{x}_n\in\mathbb{R}^d\right\}^N_{n=1}$. A linear classifier is given in terms of the inner product:
\begin{equation}
\label{eq:linearInner}
    f(\bm{x})=\left<\bm{w},\bm{x}\right>+w_0.
\end{equation}

\noindent In case of the linear SVM, when training dataset is represented by support vectors $\{\bm{x}_i\}^K_{i=1}$, $K\ll N$ we have:

\begin{equation}
\label{eq:svm}
    \bm{w}=\sum^K_{i=1}\alpha_i\bm{x}_i
\end{equation}

\noindent where $\{\alpha_i\}^K_{i=1}$ are dual variables learned from training data. A nonlinear SVM is given in terms of the kernel functions:

\begin{equation}
\label{eq:nonlinearSVM}
    f(\bm{x})=\sum^K_{i=1}\beta_i\kappa(\bm{x},\bm{x}_i)+\beta_0.
\end{equation}

\noindent By invoking the kernel trick, we can write (\ref{eq:nonlinearSVM}) as:

\begin{equation}
\label{eq:kernelTrick}
    f(\bm{x})=\sum^K_{i=1}\beta_i\left < \phi(\bm{x}_i),\phi(\bm{x})\right>_\mathcal{H}+w_0 = \left< \sum^K_{i=1}\beta_i\phi(\bm{x}_i),\phi(\bm{x})\right>_\mathcal{H}+w_0 = \left< \bm{w},\phi(\bm{x})\right>_\mathcal{H}+w_0.
\end{equation}

\noindent which is structurally equivalent to (\ref{eq:linearInner}), i.e. $\bm{w}=\sum^K_{i=1}\beta_i\phi(\bm{x}_i)$. Evaluation of one kernel function in (\ref{eq:nonlinearSVM}) is computationally proportional to evaluation of inner product in (\ref{eq:linearInner}). Thus, kernel-based nonlinear algorithm is computationally $K$ times more demanding than linear algorithms. On the other hand, direct evaluation of (\ref{eq:kernelTrick}) requires only one inner product, but feature map $\phi(\bm{x})$ can be infinite dimensional. However, if $d$ is small and $\phi(\bm{x})$ is approximated by $\phi_m(\bm{x})$, dimension of $\mathcal{H}_D$ given by (\ref{eq:Choose}) is not too large (see Appendix A for details). Thus, direct evaluation of (\ref{eq:kernelTrick}) becomes computationally feasible. Hence, nonlinear SVM or logistic calssifiers can be implemented through:

\begin{equation}
\label{eq:svmLogisticImpl}
    f(\bm{x})=\left< \bm{w}_m,\phi_m(\bm{x})\right>_\mathcal{H}+w_0.
\end{equation}

\noindent where linear classifier in feature space, represented in terms of $(\bm{w}_m,w_0)$ is learned on the training set by minimizing the appropriate loss function. The number of coefficients to be learned for linear SVM classifier and logistic classifier in the input space is $d+1$, and in the aEFM-induced space it is $D+1$. For $d=3$ (e.q. RGB feature) and $m=2$ that yields 11, and for $m=3$ it yields 21 learnable parameter (see Eq. (\ref{eq:Choose})). Thus, the increase in complexity is modest, considering the fact that linear classifiers in aEFM-induced space yield improved performance, remain interpretable and through incremental learning mode can address very large-scale classification problems. Creation of incremental learning models for nonlinear kernel-based algorithms requires some type of linearization \citep{gepperth2016incremental}. Thus, direct implementation of linear models in aEFM-induced space appears as a logical choice. Proposed approach is naturally extendable to ensemble learning. For example, we can train classifier in input or aEFM-induced space on the subsets of the training set. In the application considered in this paper, each image in the training set was used as a subset. Afterwards, final decision is reached by using majority vote to combine results of the trained classifiers achieved on the test set \citep{polikar2001learn,wen2007incremental}. As shown in section \ref{sec:experimental}, such approach significantly improves the performance of algorithms in the input space and modestly of the algorithms in the aEFM-induced space.

\subsection{Clustering and explicit feature maps}
\label{subsec:clustAndExplicit}
We want to demonstrate that, besides classification, use of EFMs can also improve performance of clustering of low-dimensional datasets. Thus, we apply the U-SPEC clustering algorithm \citep{huang2020ultra} in the aEFM-induced space. Here, image segmentation is considered as a problem of clustering pixels into cancerous and noncancerous groups. Because of their ability to handle non-linearly separable datasets, spectral clustering algorithms have gained significant attention in recent years \citep{ng2001spectral,von2007tutorial,cai2015large,he2019fast}. This statement applies in particular to subspace clustering algorithms that explore topological pair-wise relations of data with graph structure \citep{elhamifar2013sparse,liu2013robust,brbic2020ell_0}. These algorithms construct Laplacian matrix based on estimated affinity matrix and apply k-means algorithm to eigen-vectors of the Laplacian matrix to obtain clustering results \citep{von2007tutorial}. However, it takes $O(N^2d)$ time and $O(N^2)$ memory to construct the affinity matrix. Time and memory complexities for solving the eigen-decomposition problem \citep{huang2020ultra,he2019fast} are respectively $O(N^3)$ time and $O(N^2)$. Obviously, such computational complexity hinders applicability of spectral clustering algorithms to large-scale clustering problems, such as pixel wise semantic segmentation of histopathological images of $H\&E$ stained frozen sections. 

To address such a challenging problem, the U-SPEC algorithm replaces $N\times N$ graph construction problem with the $N\times p$ bi-partite graph construction problem, where $p\ll N$ stands for the number of representatives (a.k.a. anchors or landmarks). The time and memory complexity of the U-SPEC algorithm is $O(Np^\frac{1}{2}d)$ and $O(Np^\frac{1}{2})$, respectively. The method is characterized by three hyper-parameters: $c$-number of clusters, $k$-number of neighbors necessary to build sparse cross-affinity matrix of bi-partite graph, and $p$-number of representatives. As shown in section \ref{sec:experimental}, the U-SPEC method was applied successfully to semantic segmentation of original and aEFM mapped images of $H\&E$ stained frozen sections. The MATLAB code of the U-SPEC algorithm is available at \citep{huang2020ultraCode}.

\subsection{Interpretability and explainability}
In the medical sector, explanation for decisions of machine learning algorithms is required for the sake of accountability and transparency \citep{rudin2019stop,tjoa2021asurvey}. It is therefore recommended to focus on creating interpretable models \citep{rudin2019stop}. We show that, when features in the original input space are interpretable, linear classification methods remain interpretable in aEFM-induced space, too. We thoroughly illustrate the proposed methodology on images contained in the CoCaHis dataset \citep{cocahis2021}. There, each data point is a pixel in 3D space, i.e. dimension of the input space is $d=3$. Features $x_1$, $x_2$, and $x_3$ represent red, green, and blue color components, respectively. By using expression (\ref{eq:appx15}) in Appendix A, the EFM $\phi_2(\bm{x})$ associated with the polynomial kernel of order $m=2$ and $b=1$, is given in analytical form as:

\begin{equation}
\label{eq:polyAnalytical}
    \phi_2(\bm{x})=\left[{1}\ {\frac{x_1}{\sqrt{2}}}\  {\frac{x_2}{\sqrt{2}}}\ {\frac{x_3}{\sqrt{2}}}\ {\sqrt{2}x_1x_2}\ {\sqrt{2} x_1 x_3}\ {\sqrt{2} x_2 x_3}\ x^2_1\ x^2_2\ x^2_3\right]^T.
\end{equation}

\noindent aEFM associated with the Gaussian kernel of order $m=2$ and $s>0$ has, in terms of the feature monomials, the same structure as (\ref{eq:polyAnalytical}) but different coefficients. Thus, if we can explain contributions of individual features towards prediction in the input space, we should be able to explain them in the aEFM-induced space as well. We substantiate that by estimating the Shapely additive explanation values \citep{lundberg2017unified,aas2021explaining} in the input space and aEFM-induced space. By them, features contributions towards individual predictions of the logistic classifier can be explained and quantified (see section \ref{subsec:interpretExplan}).

For a training set $\{y_i,\bm{x}_i \}^{N_{train}}_{i=1}$ of the size $N_{train}$, we introduce Shapely additive explanation values $\{a_j \}^d_{j=0}$ for a  predictive model $f(\bm{x}^*)$ in the input space as \citep{lundberg2017unified,aas2021explaining}:

\begin{equation}
\label{eq:shapely}
    f(\bm{x}^*)=a_0+\sum^d_{j=1}a_j
\end{equation} 

\noindent where $a_0=E(f(\bm{x}))$. Hence by summing up the Shapely values for different features, we obtain the feature efficiency:

\begin{equation}
\label{eq:efficiency}
    \mathrm{efficiency}=\sum^d_{j=1}a_j = f(\bm{x}^*) - E\left[ f(\bm{x}) \right]
\end{equation}

\noindent Feature efficiency shows us the part of the prediction value that is not explained by the global mean prediction but it is explained by the features (Aas et al., 2021). We are also going to use average feature contributions $\{\bar{a}_j\}^d_{j=1}$ obtained as a mean of the values of each contribution across all the data in the training set, i.e.:

\begin{equation}
\label{eq:avgContrib}
    \bar{a}_j=\frac{1}{N_{train}} \sum^{N_{train}}_{n=1}a_j(n), j=1,\dots,d.
\end{equation}

\noindent Average contribution (\ref{eq:avgContrib}) indicates how much each feature deviated from the globally expected value across all predictions. It is straightforward to extend (\ref{eq:shapely}), (\ref{eq:efficiency}), and (\ref{eq:avgContrib}) to features in the aEFM induced space for a predictive model. Thus, we can quantify contribution of the aEFM to the prediction performance.

\section{Experimental results}
\label{sec:experimental}

We demonstrate proposed methodology on a very large-scale problem related to intraoperative pixel-wise classification (diagnosis) and clustering of adenocarcinoma of a colon in a liver \citep{sitnik2021cocahis}. For classification, we used linear SVM and linear logistic classifier in input space and aEFM-induced space. Also, for clustering, the U-SPEC algorithm was in the mentioned spaces.

\subsection{The CoCaHis dataset}
\label{subsec:cocahis}
Publicly available CoCaHis dataset \citep{cocahis2021} is described in details in \citep{sitnik2021cocahis}. In brief, the dataset contains 82 images of $H\&E$ stained frozen sections collected intraoperatively from 19 patients diagnosed with the adenocarcinoma of a colon in a liver. Each image is of the size $ 1388\times 1037= 1,439,456 $ pixels. 58 images (~70\%) belonging to 13 patients (~70\%) were used for training or hyper-parameter selection. Further, 24 of images (~30\%) corresponding to 6 patients (~30\%) were used for testing. Seven experts (four pathologists, two residents of pathology and one final-year student of medicine) performed pixel wise labeling. Majority vote was used to obtain the final ground truth necessary for training, hyper parameter selection and performance evaluation. The Fleiss’ kappa statistic \citep{fleiss1971measuring} of 0.74 indicated substantial inter-annotator agreement \citep{landis1977measurement}. As it is seen, the training set comprises 83,482,648 pixels. Thus, kernel-based nonlinear classifiers training in a batch mode is impossible on such a large dataset. On the contrary, implementation of linear classifiers in the aEFM-induced space in the incremental learning mode is straightforward.

\subsection{Clustering and classification algorithms}
\label{subsec:clustAndClass}
All used algorithms were implemented in the Matlab script language version 2021a. The U-SPEC algorithm was available at \citep{huang2020ultraCode}. The linear SVM algorithm was implemented using the \textit{fitcsvm} function with the iterative single data algorithm \citep{huang2006kernel} on standardized data. The logistic classifier was implemented using the \textit{fitclinear}. Afterwards, the incremental learning through the \textit{incrementalLearner} function was developed for both classifiers. Incremental learning was executed on patches of the size of $100 \times 100$ pixels, whereat the learner was implemented by the scale-invariant solver \citep{kempka2019adaptive}.

\subsection{Train and test protocols for clustering and classification}
\label{subsec:trTstProtocols}

The U-SPEC algorithm performs clustering, i.e. it is an unsupervised method and does not need labels for training. However, it is characterized by the three hyper-parameters: number of clusters $c$, number of anchors $p$ (necessary for construction of bi-partite graph), and number of the nearest neighbors $k$ (needed for similarity matrix construction). We used the training set to select the optimal triplet ($c^*$, $p^*$, $k^*$). Afterwards, we applied the U-SPEC algorithm to images from the test set with the selected hyper-parameters. The optimal triplet was selected from $c\in \{2, 3, 4, 5, 6\}$, $p\in \{75, 125, 250, 500, 1000, 2000, 4000, 8000\}$, and $k\in\{3, 5, 7\}$. Obtained optimal values were $c^*=2$, $p^*=75$ and $k^*=3$. Although we are interested in binary image segmentation, the non-cancerous class is heterogeneous and several types of the tissues can be present in the specimen \citep{kopriva2015unsupervised}. That justifies our decision to select the number of clusters from the set specified above. The hyper-parameters were selected based on training data set in the RGB color space. They were used for semantic segmentation of test set images in the input RGB color space as well as in the aEFM-induced spaces. 

As explained in section \ref{subsec:clustAndClass}, the U-SPEC algorithm was applied on an image-by-image basis to address the large-scale clustering problem. To account for the noise presence, results obtained by the U-SPEC algorithm were filtered by $9\times 9$ 2D median filter implemented by function \textit{medfilt2}. The linear SVM classifier and linear logistic classifier were trained in the incremental learning mode for each type of data. Afterwards, the trained models were applied to unseen test data using function \textit{predict}. Original images in the RGB color space were scaled to [0, 1] interval prior to any processing. 

For classification problem, original images in the RGB color space were denoised by a function \textit{wdenoise2} that implements 2D wavelet transform. The denoising was performed on each color channel separately. Regarding ensemble mode, linear classifiers in the input space and aEFM-induced space were trained on 58 images from the training set, and applied to test images. The final selection of models was obtained through the majority vote. For each training image, a classifier associated with the polynomial EFM was selected from the candidates with offsets $b\in\{1, 2, 3, 4, 5, 6, 7\}$. Criterion for selection was the maximum balanced accuracy. The same approach was used for classifiers associated with the Gaussian aEFM for standard deviation $\sigma \in\{4, 2\sqrt{2}, 2, \sqrt{2}, 1, 1/\sqrt{2}, 1/2, \sqrt{2}/4\}$.

\subsection{Performance measures}

We used the following micro measures for validating models' performances:
\begin{itemize}

\item sensitivity (SE), a.k.a. recall and true positive rate
\begin{equation}
\label{eq:sens}
    SE = \frac{TP}{TP+FN},
\end{equation}

\item specificity (SP), a.k.a. selectivity and true negative rate
\begin{equation}
\label{eq:spec}
    SP = \frac{TN}{{TN}+{FP}},
\end{equation}

\item $\mathrm{F}_1$ score, a.k.a. Dice coefficient
\begin{equation}
\label{eq:f1}
    F_1 = \frac{2TP}{2TP+FP+FN},
\end{equation}

\item positive predicted value (PPV), a.k.a. precision
\begin{equation}
\label{eq:ppv}
    PPV = \frac{TP}{TP+FP},
\end{equation}

\item balanced accuracy (BACC)
\begin{equation}
\label{eq:bacc}
    BACC = \frac{SE+SP}{2}.
\end{equation}
\end{itemize}

\noindent These metrics are defined in terms of true positives (TP), correctly diagnosed cancerous pixels; true negatives (TN), correctly diagnosed non-cancerous pixels; false positives (FP), incorrectly diagnosed non-cancerous pixels, and false negatives (FN) incorrectly diagnosed cancerous pixels. Micro performance implies that TP, TN, FP and FN were estimated cumulatively for all the images in the test set. For all the metrics, value 0 represents the worse result and value 1 indicates the best result.

\subsection{Interpretation and explanation of diagnosis of adenocarcinoma of a colon in a liver}
\label{subsec:interpretExplan}
\begin{figure}[!ht]
\centering
\includegraphics[width=0.8\textwidth]{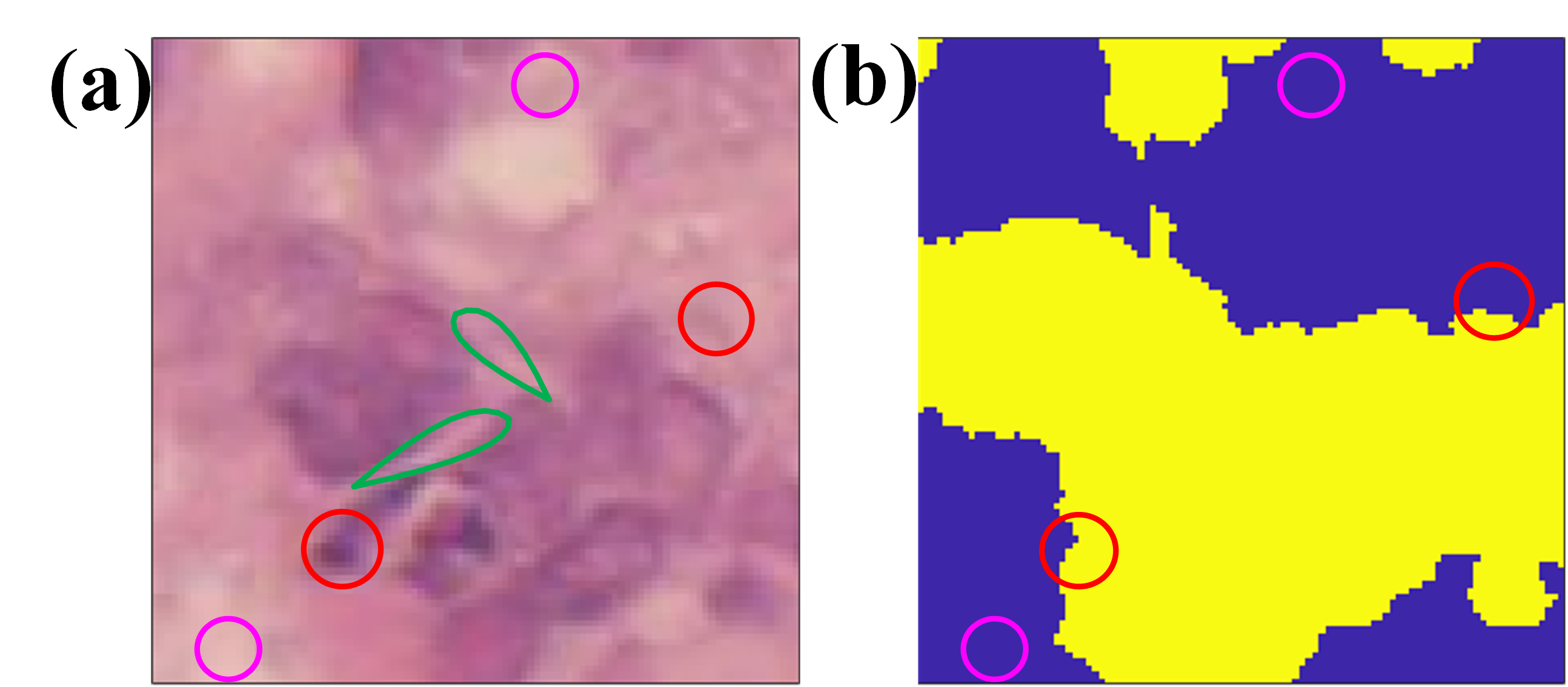}
\caption{Image (a) represents $100\times 100$ pixels patch of the $H\&E$ stained image of frozen section with adenocarcinoma of a colon in a liver. The ground truth obtained by majority vote of pixel wise annotations by seven experts can be seen in (b). Yellow color indicates cancerous pixels and circles indicates different regions of interest.}
\label{fig:patch} 
\end{figure}

\begin{figure}[!ht]
    \centering
    \includegraphics[width=0.9\textwidth]{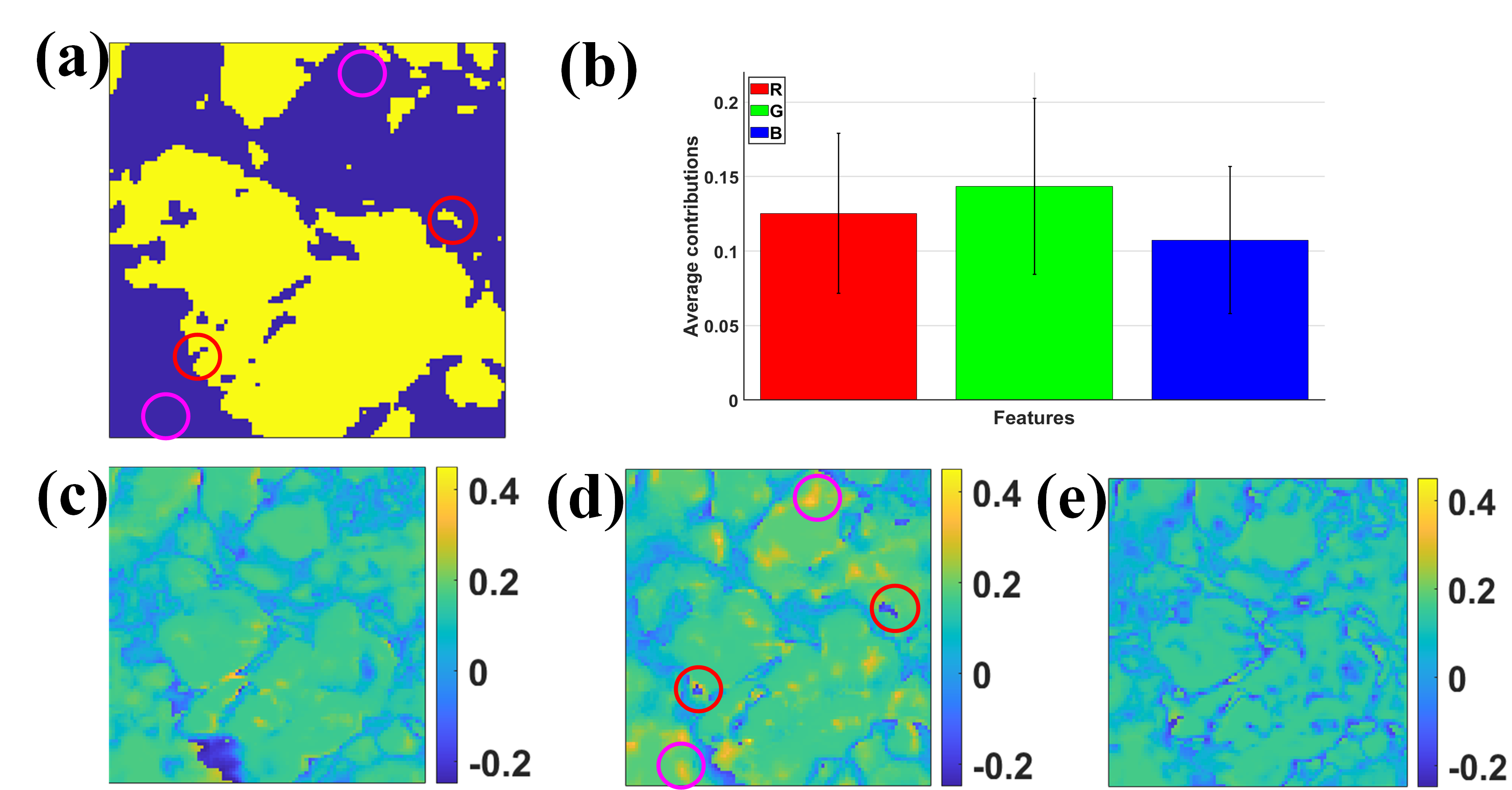}
     \caption{Interpretability and explainability of predictions for a patch shown in Fig. \ref{fig:patch}(a) and corresponding ground truth shown in Fig. \ref{fig:patch}(b). (a) represents labels predicted by the logistic classifier trained in the input RGB color space. Yellow color indicates cancerous pixels. (b) averaged contributions of R, G and B features in explaining model predictions. Estimated Shapely additive explanation values for individual pixel-wise predictions: (c) $R$ feature, (d) $G$ feature, (e) $B$ feature. Circles represent regions of interest.}
    \label{fig:intExpl}
\end{figure}

We selected a $100\times 100$ pixels patch to show interpretability, explainability and feature efficiency in terms of Shapely additive values. The patch is shown in Fig. \ref{fig:patch}(a) with the ground truth labels shown in Fig. \ref{fig:patch}(b). The patch contains 52\% cancerous pixels. To make the presentation easier to follow, we name the input space features $x_1$, $x_2$, and $x_3$ respectively as red (R), green (G) and blue (B). Fig. \ref{fig:intExpl}(a) shows labels predicted by the logistic classifier in the input RGB color space. According to (\ref{eq:avgContrib}), the average feature contributions were estimated and shown in Fig. \ref{fig:intExpl}(b), while pixel-wise individual contributions are shown in Figs. \ref{fig:intExpl}(c) to \ref{fig:intExpl}(e). We used MATLAB function \textit{shapely} with the extended Kernel SHAP algorithm to estimate Shapely additive values for individual prediction made by the logistic classifier. Estimation of the contributions of individual features is a demanding problem when features are dependent (correlated). Because of the tissue coloring, that is the case with the $R$, $G$ and $B$ features in the considered problem. Thus, when using the Kernel SHAP algorithm, we selected the empirical estimate of the conditional distribution for the evaluation of the value function \citep{aas2021explaining}. A large training set with dependent features is why it becomes highly demanding to computationally estimate Shapely additive values for individual predictions. Demonstration of interpretability and explainability was performed on a patch of the size of $100\times 100$ pixels. 

Furthermore, estimated Shapely additive values enabled us to calculate feature efficiency in accordance with (9). Thus, we present in Table \ref{tab:logisticClsPerformance} results obtained by the logistic classifier trained in the input space and in the space induced by polynomial EFMs of the order $m\in\{2, 3, 4\}$ and with $b=1$. Results are presented in terms of BACC, $\mathrm{F}_1$ score, PPV and feature efficiency. When compared to the efficiency in the input space, feature efficiency in the space induced by the polynomial EFMs has been increased in absolute values by 2.69\%, 3.78\% and 4.31\% for $m=2$, 3, and 4, respectively. Thus, improved performance of logistic classifier in EFM induced space is due to the EFM mapping, i.e. because of the newly constructed features that do not exist in the input RGB color space.

\begin{table}
\centering
\caption{Performance of linear logistic classifier applied to a patch of the size $100\times 100$ pixels in the input RGB color space and in the space induced by the polynomial EFMs of order $m\in\{2, 3, 4\}$ with $b=1$. Original images were denoised using MATLAB function \textit{wdenoise2} prior to further processing.}
\label{tab:logisticClsPerformance}
\begin{tabular}{|l|c|c|c|c|}
\hline
{\color[HTML]{3D3D3D} } &
  {\color[HTML]{3D3D3D} \textbf{BACC}} &
  {\color[HTML]{3D3D3D} \textbf{F1 score}} &
  {\color[HTML]{3D3D3D} \textbf{PPV}} &
  {\color[HTML]{3D3D3D} \textbf{Feature efficiency}} \\ \hline
{\color[HTML]{3D3D3D} Input space} &
  {\color[HTML]{3D3D3D} 0.8808} &
  {\color[HTML]{3D3D3D} 0.8827} &
  {\color[HTML]{3D3D3D} 0.9049} &
  {\color[HTML]{3D3D3D} 0.3450} \\ \hline
{\color[HTML]{3D3D3D} \begin{tabular}[c]{@{}l@{}}poly EFM \\ \{m=2, b=1\}\end{tabular}} &
  {\color[HTML]{3D3D3D} 0.8889} &
  {\color[HTML]{3D3D3D} 0.8918} &
  {\color[HTML]{3D3D3D} 0.9057} &
  {\color[HTML]{3D3D3D} 0.3712} \\ \hline
{\color[HTML]{3D3D3D} \begin{tabular}[c]{@{}l@{}}poly EFM \\ \{m=3, b=1\}\end{tabular}} &
  {\color[HTML]{3D3D3D} 0.8947} &
  {\color[HTML]{3D3D3D} 0.8983} &
  {\color[HTML]{3D3D3D} 0.9066} &
  {\color[HTML]{3D3D3D} 0.3828} \\ \hline
{\color[HTML]{3D3D3D} \begin{tabular}[c]{@{}l@{}}poly EFM \\ \{m=4, b=1\}\end{tabular}} &
  {\color[HTML]{3D3D3D} \textbf{0.8965}} &
  {\color[HTML]{3D3D3D} \textbf{0.9001}} &
  {\color[HTML]{3D3D3D} \textbf{0.9076}} &
  {\color[HTML]{3D3D3D} \textbf{0.3881}} \\ \hline
\end{tabular}
\end{table}

We now want to interpret contributions of each feature in terms of medical meaning. In images of $H\&E$ stained specimens, nucleus, cytoplasm and glandular structures appear respectively blue-purple, pink and white. Intensely red structures may also represent red blood cells \citep{kothari2014removing}. From color formation theory, we know that red and blue colors dominate in formation of the purple color. The pink color is a mix of red and white color, while the white color itself is a mix of red, green and blue colors. Hence, by using color formation theory and combining Figs. \ref{fig:patch}(a) and \ref{fig:patch}(b) it is justified to expect that three colors exhibit positive correlations with the prediction probability of cancer class, as well as with the noncancerous class. That is visible in Figs. \ref{fig:intExpl}(c) to \ref{fig:intExpl}(e). Positive correlation implies that increase in the feature value will increase the probability of corresponding outcome. Interpretation of negative correlation is the opposite. Thus, by manipulating values of the $R$, $G$ and $B$ features, the classifier can simultaneously increase sensitivity and specificity. In that regard, Fig. \ref{fig:intExpl}(b) further confirms that, on average, all three features have a positive impact on the prediction probabilities of the classifier decision. 

It is seen in Fig. \ref{fig:intExpl}(d) that the existence of dark blue spots (encircled red in related panels) where the green color exhibits negative correlation with the prediction probability of cancer class. That agrees with the predicted values in Fig. \ref{fig:intExpl}(a), where small related areas are colored yellow (cancer class) or blue (non-cancer class). It is, however, not in agreement with the annotations shown in Fig. \ref{fig:patch}(b) and with the Fig. \ref{fig:patch}(a). The upper red circle is blue and should be, according to color formation theory, annotated as cancer. However, it is annotated as noncancerous in Fig. \ref{fig:patch}(b) and there is a possibility that expert labeling is unprecise. Thus, Shapely additive value helped us to spot possibly incorrectly annotated regions.

Furthermore, there are yellow spots (encircled in magenta color in related panels) in Fig. \ref{fig:intExpl}(d) where positive correlations of the green color are significant. According to Fig. \ref{fig:patch}(b), these spots are in cancerous and non-cancerous regions and have to be related to color of the corresponding positions in Fig. \ref{fig:patch}(a). As an example, the lower left corner is colored white and, therefore, an increase in the green value will increase probability of prediction of the noncancerous class. The same is true for the upper middle part in Fig. \ref{fig:patch}(a). In Fig. \ref{fig:intExpl}(c) we see that red color exhibit more positive correlations in the cancerous region, and more negative correlations in the noncancerous region. That agrees with the color formation theory, because red color is less present in locations of pink and white colors that occupy the noncancerous region of the image. Regarding the blue color feature, Fig. \ref{fig:intExpl}(e) shows even more expressed negative correlations in the noncancerous region. That is in agreement with the color formation theory because the blue color is the least important for the formation of the pink and white colors, which appear dominantly to the noncancerous region. 

It is also seen in Figs. \ref{fig:intExpl}(c) to \ref{fig:intExpl}(d) that positive correlations with the probability of the cancer class are not spatially homogeneous. The cancerous region is intersected with the lines of negative correlations. These lines coincide with the narrow white lines (encircled green) in Fig. \ref{fig:patch}(a), which indicate possibly noncancerous pixels. Thus, experts’ based labeling shown in Fig. \ref{fig:patch}(b) may be incorrect in these areas.

\begin{figure}[!ht]
    \centering
    \includegraphics[width=0.9\textwidth]{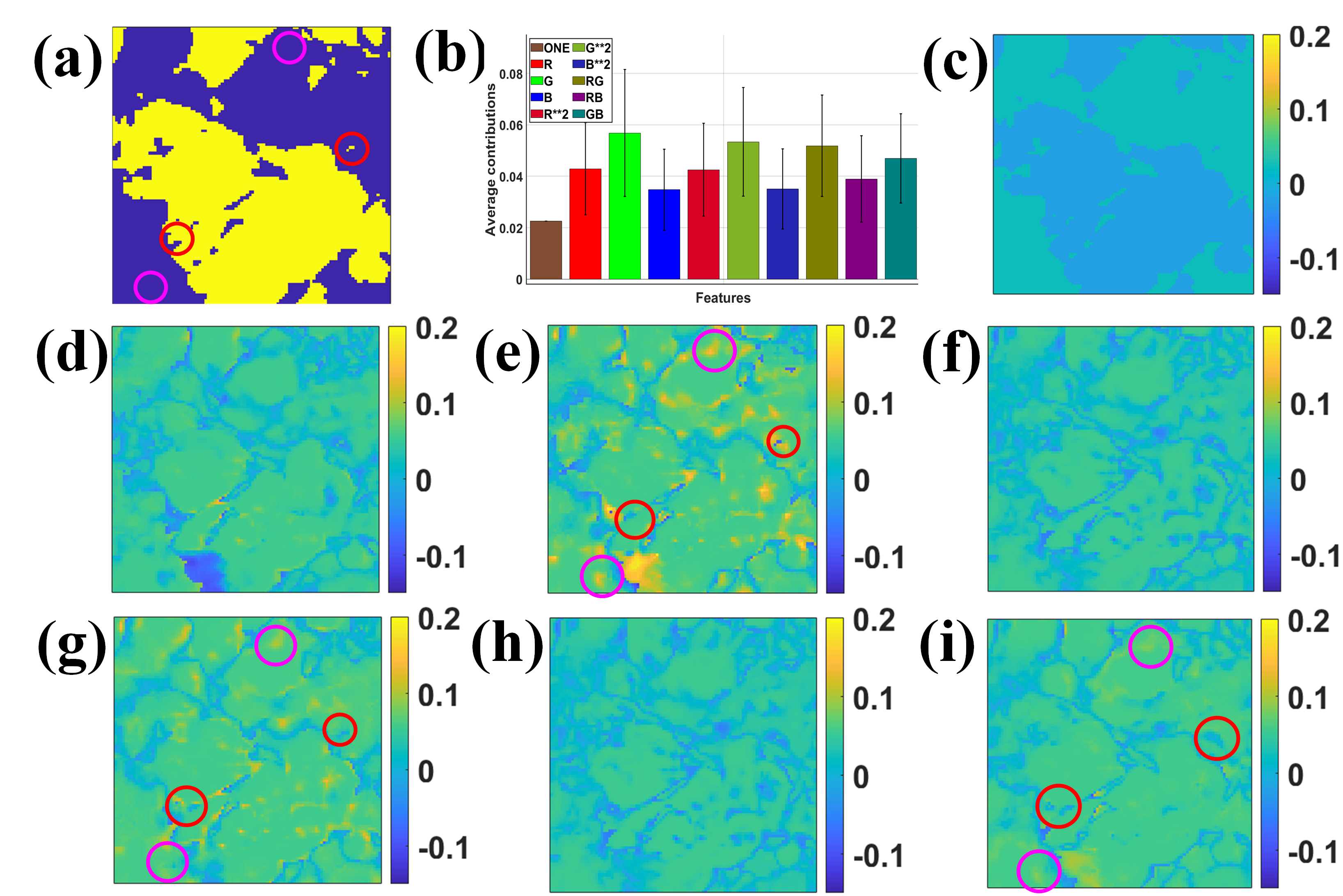}
     \caption{Interpretability and explainability of predictions for a patch shown in Fig. \ref{fig:patch}(a) with a ground truth shown in Fig. \ref{fig:patch}(b). (a) labels predicted by the logistic classifier trained in the space induced by polynomial EFM with $m=2$ and $b=1$. Yellow color indicates cancerous pixels. (b) averaged contributions of features $ONE$, $R$, $G$, $B$, $R^2$, $G^2$, $B^2$, $RG$, $RB$ and $GB$, in explaining model predictions. (c) to (i): estimated Shapely additive explanation values for individual pixel-wise predictions for features in respective order $ONE$, $R$, $G$, $B$, $RG$, $RB$ and $GB$. }
    \label{fig:intExplExpand}
\end{figure}

We now want to interpret contributions of each feature towards predictions made by a logistic classifier in a space induced by the polynomial EFM of the order $m=2$, and the offset $b=1$. Fig. \ref{fig:intExplExpand}(a) shows labels predicted by logistic classifier trained on data in the mentioned space. Contributions of each feature averaged across all the pixels, see Eq. (\ref{eq:avgContrib}), are shown in Fig. \ref{fig:intExplExpand}(b). Pixel wise individual contributions are shown in Figs. \ref{fig:intExplExpand}(c) to \ref{fig:intExplExpand}(i). Because the visual appearances of $R^2$, $G^2$ and $B^2$ features are the same as of $R$, $G$ and $B$ features, their individual contributions are not displayed. Shapely additive explanation values for $R$, $G$ and $B$ features, that are shown respectively in Figs. \ref{fig:intExplExpand}(c) to \ref{fig:intExplExpand}(e), are equivalent to the ones in the input RGB color space shown respectively in Figs. \ref{fig:intExpl}(c) to \ref{fig:intExpl}(e). Thus, their interpretation remains the same. 

We see in Fig. \ref{fig:intExplExpand}(b) that $RG$ and $GB$ features, visualized in Figs. \ref{fig:intExplExpand}(g) and \ref{fig:intExplExpand}(i), play an important role in an explanation of the model predictions in EFM induced space. Individual contributions of $RG$ feature, are comparable with the ones of $R$ and $G$ features. However, in locations of yellow color spots (encircled in magenta color) the correlations of $RG$ feature are less positive than with the $G$ feature. Also, in locations of dark blue color spots (encircled in red color) correlations of $RG$ feature are less negative than with the $R$ feature. The fact that the $RG$ and $GB$ features have, on average, a stronger impact on model predictions than the $R$ and $B$ features implies that mixtures of the basic $R$, $G$ and $B$ colors are necessary to provide better explanations of model predictions. That is reasonable because colors of cancerous and noncancerous regions are mixtures of the three basic colors. Thus, the new features that do not exist in the input RGB color (feature) space provided an improved performance of the logistic classifier in the EFM induced space. That is also confirmed by feature efficiency presented in Table \ref{tab:logisticClsPerformance}. 

Furthermore, Figs. \ref{fig:intExplExpand}(c) to \ref{fig:intExplExpand}(i) indicate that related feature contributions in the cancerous region are not spatially homogeneous. Instead, they are intersected with the lines where contributions are below the globally expected values. These lines coincide with the narrow white lines (encircled green) in Fig. \ref{fig:patch}(a), which indicate possibly noncancerous pixels. Thus, as mentioned earlier, experts’ based labeling shown in Fig. \ref{fig:patch}(b) may be unprecise in these areas. To be even more convincible in terms of interpretability of the model predictions in EFM induced space, we further comment contributions towards individual predictions of the feature $ONE$, see Eq. (\ref{eq:polyAnalytical}), shown in Fig. \ref{fig:intExplExpand}(c). Regarding the visual appearance, it completely mimics the map with the predicted labels shown in Fig. \ref{fig:intExplExpand}(a). In the cancerous region it is negatively correlated with the prediction of cancerous class. Moreover, in the noncancerous region it is positively correlated with the prediction of noncancerous class. The correlations are of the same magnitude but of the opposite sign. Since it is not a function of color, the constant feature is spatially invariant. Hence, its average contribution towards model predictions is neutral. It means that increasing the constant value will increase specificity and decrease sensitivity and vice versa. Thus, its manipulation cannot improve classifier’s performance. 

As seen in section \ref{sec:methods}, proposed approach to nonlinear classification is extendable to ensemble learning, i.e. models can be learned on subsets of the training set. Since the final prediction of the ensemble is obtained by combining predictions of individual models, it is clear that prediction obtained by ensemble learning remains interpretable as long as individual predictions are interpretable.

\subsection{Performance of clustering and classification in the input RGB color space and EFM-induced space}

We applied proposed aEFM-based concept to intraoperative semantic segmentation and classification of adenocarcinoma of a colon in a liver. To estimate statistical significance of the results in terms of performance measures (\ref{eq:sens}) to (\ref{eq:bacc}), we trained each model 40 times. To estimate performance of the ensemble of classifiers, we used 10 runs. Also, the two-samples Student’s t-test, implemented with the Matlab function \textit{ttest2}, was applied. The goal was to find out whether particular performance measure, achieved by the selected algorithm in the aEFM-induced space, is statistically significantly better than the performance of the corresponding algorithm in the input RGB color space.

Table \ref{tab:clusteringPerformance} reports the best clustering results achieved by the U-SPEC algorithm. The algorithm was applied to data in the space induced by the polynomial EFM with the parameters: $m\in\{2, 3, 4\}$ and $b\in\{0, 1, 2, 3\}$. The algorithm was applied to data in the space induced by the Gaussian aEFM with the parameters: $m\in\{2, 3\}$ and $\sigma \in\{4, 2\sqrt{2}, 2, \sqrt{2}, 1, 1/\sqrt{2}, 1/2, \sqrt{2}/4\}$. Since, results obtained in space induced by Gaussian aEFM were not better than the ones in the input space, they are not shown in Table \ref{tab:clusteringPerformance}. We also report result achieved on the same training and test datasets \citep{sitnik2021cocahis} by DeepLabV3+ algorithm \citep{chen2017deeplab,chen2017rethinking}. Achieved performance of DeepLabv3+ is significantly better but complexity of the model is huge compared to the model in EFM mapped space: $\sim$12.6 millions of parameters with the MobileNetV2 \citep{sandler2018mobile} backbone vs. 5 hyper-parameters (3 of the U-SPEC algorithm and 2 of the map). In other words, there are no learnable parameters in the U-SPEC algorithm (the $N\times p$ bi-partite graph is calculated directly from data). Thus, because there is a limited quantity of annotated data, proposed method in aEFM-induced space is expected to be more robust to overfitting.

The equivalent comment applies to U-Net \citep{ronneberger2015u} and U-Net++ \citep{zhou2018unet++} algorithms that were used for diagnosis of adenocarcinoma in a liver \citep{sitnik2021cocahis}. Number of parameters of the U-Net model is $\sim$22.5 millions with the DenseNet201 backbone \citep{huang2017densely}, and of the U-Net++ model is $\sim$48.5 millions of parameters with the DenseNet201 backbone. When compared to the model in input RGB space, the U-SPEC algorithm in the space induced by polynomial EFM of order $d=4$ achieved statistically significant improvements in BACC, $\mathrm{F}_1$ score and PPV in the amounts of 0.79\%, 0.95\% and 0.83\%, respectively.

\begin{table}
\centering
\caption{Performance of the U-SPEC algorithm on pixel wise semantic segmentation of adenocarinoma of a colon in a liver. Results are reported in terms of mean and standard deviation estimated from 40 runs. The testing results shown in boldface are statistically significantly better (with a significance level of $\alpha=0.05$) then the result in the input RGB color space. }
\label{tab:clusteringPerformance}
\begin{tabular}{|c|c|c|c|c|c|}
\hline
{\color[HTML]{3D3D3D} \textbf{Model}} &
  {\color[HTML]{3D3D3D} \textbf{SE {[}\%{]}}} &
  {\color[HTML]{3D3D3D} \textbf{SP{[}\%{]}}} &
  {\color[HTML]{3D3D3D} \textbf{BACC {[}\%{]}}} &
  {\color[HTML]{3D3D3D} \textbf{F1 {[}\%{]}}} &
  {\color[HTML]{3D3D3D} \textbf{PPV {[}\%{]}}} \\ \hline
{\color[HTML]{3D3D3D} input RGB color space} &
  {\color[HTML]{3D3D3D} 79.50±1.07} &
  {\color[HTML]{3D3D3D} 75.24±2.69} &
  {\color[HTML]{3D3D3D} 77.37±1.27} &
  {\color[HTML]{3D3D3D} 65.18±1.72} &
  {\color[HTML]{3D3D3D} 55.29±2.50} \\ \hline
{\color[HTML]{3D3D3D} \begin{tabular}[c]{@{}c@{}}polynomial EFM \\ m=2, b=1\end{tabular}} &
  {\color[HTML]{3D3D3D} 80.03±0.65} &
  {\color[HTML]{3D3D3D} 75.95±1.24} &
  {\color[HTML]{3D3D3D} 77.99±0.66} &
  {\color[HTML]{3D3D3D} 65.98±0.88} &
  {\color[HTML]{3D3D3D} 56.07±1.22} \\ \hline
{\color[HTML]{3D3D3D} \begin{tabular}[c]{@{}c@{}}polynomial EFM \\ m=3, b=1\end{tabular}} &
  {\color[HTML]{3D3D3D} 80.19±0.84} &
  {\color[HTML]{3D3D3D} 75.75±1.39} &
  {\color[HTML]{3D3D3D} 77.97±0.66} &
  {\color[HTML]{3D3D3D} 65.88±0.90} &
  {\color[HTML]{3D3D3D} 55.92±1.33} \\ \hline
{\color[HTML]{3D3D3D} \begin{tabular}[c]{@{}c@{}}polynomial EFM \\ m=4, b=1\end{tabular}} &
  {\color[HTML]{3D3D3D} \textbf{80.33±1.13}} &
  {\color[HTML]{3D3D3D} \textbf{76.00±1.06}} &
  {\color[HTML]{3D3D3D} \textbf{78.16±0.68}} &
  {\color[HTML]{3D3D3D} \textbf{66.13±0.95}} &
  {\color[HTML]{3D3D3D} \textbf{56.12±1.06}} \\ \hline
{\color[HTML]{3D3D3D} DeepLabV3+} &
  {\color[HTML]{3D3D3D} 88.24} &
  {\color[HTML]{3D3D3D} 88.79} &
  {\color[HTML]{3D3D3D} 88.52} &
  {\color[HTML]{3D3D3D} 81.14} &
  {\color[HTML]{3D3D3D} 75.10} \\ \hline
\end{tabular}
\end{table}

We report in Table \ref{tab:logisticClsPerformanceCoCaHis} the best classification results achieved by the logistic classifier on the CoCaHis test set. Trained models are validated in the input RGB color space and spaces induced by polynomial and Gaussian aEFMs. Hyper-parameters related to aEFMs were selected from the same sets used for the U-SPEC algorithm. While linear classifier in the input space includes 4 learnable parameters (coefficients), the same model, in space induced with the mapping of order $m=2$, includes 11 trainable parameters. The number of trainable parameters in space induced with mapping of order $m=3$ is 21. These complexities are highly modest when compared with the complexities of deep networks reported previously. Besides increased robustness to overfitting, linear logistic classifier in EFM induced space remains interpretable, see sections \ref{subsec:interpretExplan} and \ref{subsec:trTstProtocols}. Compared to the results in the input RGB color space, linear logistic classifier in the space induced by Gaussian aEFM of order $m=2$ achieved statistically significant absolute improvements in BACC, $\mathrm{F}_1$ score and PPV in the amounts of 12.04\%, 12.58\% and 18.17\%. 

\begin{table}
\centering
\caption{Performance of the single logistic classifier on intraoperative pixel wise classification of adenocarcinoma of a colon in a liver for CoCaHis test data in the input RGB color space and space induced by aEFMs. Results are reported in terms of mean and standard deviation estimated from 40 runs. The testing results shown in boldface are statistically significantly better (with a significance level of 0.05) than the result in the input RGB color space.}
\label{tab:logisticClsPerformanceCoCaHis}
\begin{tabular}{|c|c|c|c|c|c|}
\hline
{\color[HTML]{3D3D3D} \textbf{Model}} &
  {\color[HTML]{3D3D3D} \textbf{SE {[}\%{]}}} &
  {\color[HTML]{3D3D3D} \textbf{SP{[}\%{]}}} &
  {\color[HTML]{3D3D3D} \textbf{BACC {[}\%{]}}} &
  {\color[HTML]{3D3D3D} \textbf{F1 {[}\%{]}}} &
  {\color[HTML]{3D3D3D} \textbf{PPV {[}\%{]}}} \\ \hline
{\color[HTML]{3D3D3D} input RGB color space} &
  {\color[HTML]{3D3D3D} \textbf{94.06±0.23}} &
  {\color[HTML]{3D3D3D} 36.20±0.87} &
  {\color[HTML]{3D3D3D} 65.13±0.32} &
  {\color[HTML]{3D3D3D} 52.17±0.23} &
  {\color[HTML]{3D3D3D} 36.10±0.26} \\ \hline
{\color[HTML]{3D3D3D} \begin{tabular}[c]{@{}c@{}}polynomial EFM \\ m=3, b=1\end{tabular}} &
  {\color[HTML]{3D3D3D} 86.23±0.06} &
  {\color[HTML]{3D3D3D} 58.40±0.16} &
  {\color[HTML]{3D3D3D} 72.31±0.05} &
  {\color[HTML]{3D3D3D} 58.50±0.05} &
  {\color[HTML]{3D3D3D} 44.26±0.08} \\ \hline
{\color[HTML]{3D3D3D} \begin{tabular}[c]{@{}c@{}}Gaussian aEFM \\ m=2, $\sigma=1/2$\end{tabular}} &
  {\color[HTML]{3D3D3D} 80.26±0.12} &
  {\color[HTML]{3D3D3D} \textbf{74.09±0.14}} &
  {\color[HTML]{3D3D3D} \textbf{77.17±0.01}} &
  {\color[HTML]{3D3D3D} \textbf{64.75±0.03}} &
  {\color[HTML]{3D3D3D} \textbf{54.27±0.01}} \\ \hline
{\color[HTML]{3D3D3D} DeepLabV3+} &
  {\color[HTML]{3D3D3D} 88.24} &
  {\color[HTML]{3D3D3D} 88.79} &
  {\color[HTML]{3D3D3D} 88.52} &
  {\color[HTML]{3D3D3D} 81.14} &
  {\color[HTML]{3D3D3D} 75.10} \\ \hline
\end{tabular}
\end{table}

In Table \ref{tab:logisticEnsemblePerformanceCoCaHis} the best classification results achieved by the ensemble of the logistic classifiers are shown. It is seen that classification performance in the input RGB color space is improved significantly compared to a single classifier performance. Also, performances of the ensemble of classifiers in space induced by polynomial EFM and Gaussian aEFM are better than performance of the ensemble of classifiers in the input space. It is, however, important to notice that the best performance in terms of micro BACC is achieved by a single classifier learned in space induced by Gaussian aEFM of order $m=2$. That is important because training the ensemble of classifiers is computationally very demanding.

\begin{table}
\centering
\caption{Performance of the ensemble of 58 logistic classifiers on intraoperative pixel wise classification of adenocarcinoma of a colon in a liver for CoCaHis test data in the input RGB color space and space induced by aEFMs. Each classifier is learned on one image from the training set. Results are reported in terms of mean and standard deviation estimated from 10 runs. The testing results shown in boldface are statistically significantly better (with a significance level of 0.05) than the result in the input RGB color space. }
\label{tab:logisticEnsemblePerformanceCoCaHis}
\begin{tabular}{|c|c|c|c|c|c|}
\hline
{\color[HTML]{3D3D3D} \textbf{Model}} &
  {\color[HTML]{3D3D3D} \textbf{SE {[}\%{]}}} &
  {\color[HTML]{3D3D3D} \textbf{SP{[}\%{]}}} &
  {\color[HTML]{3D3D3D} \textbf{BACC {[}\%{]}}} &
  {\color[HTML]{3D3D3D} \textbf{F1 {[}\%{]}}} &
  {\color[HTML]{3D3D3D} \textbf{PPV {[}\%{]}}} \\ \hline
{\color[HTML]{3D3D3D} input RGB color space} &
  {\color[HTML]{3D3D3D} 72.00±0.75} &
  {\color[HTML]{3D3D3D} 79.95±0.60} &
  {\color[HTML]{3D3D3D} 75.97±0.08} &
  {\color[HTML]{3D3D3D} 64.19±0.05} &
  {\color[HTML]{3D3D3D} 57.92±0.47} \\ \hline
{\color[HTML]{3D3D3D} \begin{tabular}[c]{@{}c@{}}polynomial EFM \\ $m=3$, $b \in \{1, \dots , 7\}$ 
\end{tabular}} &
  {\color[HTML]{3D3D3D} \textbf{79.49±0.09}} &
  {\color[HTML]{3D3D3D} 74.55±0.11} &
  {\color[HTML]{3D3D3D} \textbf{77.02±0.03}} &
  {\color[HTML]{3D3D3D} 64.05±0.05} &
  {\color[HTML]{3D3D3D} 54.48±0.08} \\ \hline
{\color[HTML]{3D3D3D} \begin{tabular}[c]{@{}c@{}}Gaussian aEFM \\ $m=2$, $\sigma \in\{4, 2\sqrt{2}, 2, \sqrt{2}, 1, 1/\sqrt{2}, 1/2, \sqrt{2}/4\}$\end{tabular}} &
  {\color[HTML]{3D3D3D} 74.36±0.33} &
  {\color[HTML]{3D3D3D} 79.33±0.39} &
  {\color[HTML]{3D3D3D} 76.85±0.06} &
  {\color[HTML]{3D3D3D} \textbf{65.15±0.12}} &
  {\color[HTML]{3D3D3D} \textbf{57.96±0.47}} \\ \hline
\end{tabular}
\end{table}

In Table \ref{tab:linearSVMPerf}, we report the best classification results achieved by the linear SVM classifier in the input RGB color space and spaces induced by polynomial and Gaussian aEFMs. Hyper-parameters related to aEFMs were selected from the same sets used for the U-SPEC algorithm. Computational complexity of the linear SVM classifier is essentially the same as the one of the linear logistic classifier. Thus, the same comments apply when linear SVM is applied in the aEFM induced space. As seen in Table \ref{tab:linearSVMPerf}, compared to the results in the input RGB color space, linear SVM classifiers in the space induced by EFMs achieved statistically significant improvements in BACC, $\mathrm{F}_1$ score and PPV. The highest absolute improvements are achieved by the classifier in the space induced by the Gaussian aEFM of order $m=2$ - the amounts of 8.04\%, 9.41\% and 14.42\%, respectively. 

\begin{table}
\centering
\caption{Performance of the linear SVM classifier on intraoperative pixel wise classification of adenocarinoma of a colon in a liver for data in the input RGB color space and space induced by aEFMs. Results are reported in terms of mean and standard deviation estimated from 40 runs. The testing results shown in boldface are statistically significantly better (with a significance level of $\alpha=0.05$) than result in the input RGB color space. }
\label{tab:linearSVMPerf}
\begin{tabular}{|c|c|c|c|c|c|}
\hline
{\color[HTML]{3D3D3D} \textbf{Model}} &
  {\color[HTML]{3D3D3D} \textbf{SE {[}\%{]}}} &
  {\color[HTML]{3D3D3D} \textbf{SP{[}\%{]}}} &
  {\color[HTML]{3D3D3D} \textbf{BACC {[}\%{]}}} &
  {\color[HTML]{3D3D3D} \textbf{F1 {[}\%{]}}} &
  {\color[HTML]{3D3D3D} \textbf{PPV {[}\%{]}}} \\ \hline
{\color[HTML]{3D3D3D} input RGB color space} &
  {\color[HTML]{3D3D3D} \textbf{91.70±0.18}} &
  {\color[HTML]{3D3D3D} 46.49±0.59} &
  {\color[HTML]{3D3D3D} 69.07±0.21} &
  {\color[HTML]{3D3D3D} 55.23±0.18} &
  {\color[HTML]{3D3D3D} 39.62±0.22} \\ \hline
{\color[HTML]{3D3D3D} \begin{tabular}[c]{@{}c@{}}polynomial EFM \\ m=2, b=1\end{tabular}} &
  {\color[HTML]{3D3D3D} 89.49±0.11} &
  {\color[HTML]{3D3D3D} 54.62±0.35} &
  {\color[HTML]{3D3D3D} 71.55±0.12} &
  {\color[HTML]{3D3D3D} 57.63±0.11} &
  {\color[HTML]{3D3D3D} 42.50±0.15} \\ \hline
{\color[HTML]{3D3D3D} \begin{tabular}[c]{@{}c@{}}polynomial EFM \\ m=3, b=2\end{tabular}} &
  {\color[HTML]{3D3D3D} 85.88±0.08} &
  {\color[HTML]{3D3D3D} 62.32±0.19} &
  {\color[HTML]{3D3D3D} 73.95±0.05} &
  {\color[HTML]{3D3D3D} 60.28±0.07} &
  {\color[HTML]{3D3D3D} 46.53±0.10} \\ \hline
{\color[HTML]{3D3D3D} \begin{tabular}[c]{@{}c@{}}polynomial EM \\ m=4, b=1\end{tabular}} &
  {\color[HTML]{3D3D3D} 83.47±0.08} &
  {\color[HTML]{3D3D3D} 65.12±0.15} &
  {\color[HTML]{3D3D3D} 74.30±0.04} &
  {\color[HTML]{3D3D3D} 60.81±0.05} &
  {\color[HTML]{3D3D3D} 47.83±0.09} \\ \hline
{\color[HTML]{3D3D3D} \begin{tabular}[c]{@{}c@{}}Gaussian aEFM \\ m=2, $\sigma=1/2$\end{tabular}} &
  {\color[HTML]{3D3D3D} 80.43±0.12} &
  {\color[HTML]{3D3D3D} \textbf{73.79±0.14}} &
  {\color[HTML]{3D3D3D} \textbf{77.11±0.01}} &
  {\color[HTML]{3D3D3D} \textbf{64.64±0.03}} &
  {\color[HTML]{3D3D3D} \textbf{54.04±0.10}} \\ \hline
{\color[HTML]{3D3D3D} \begin{tabular}[c]{@{}c@{}}Gaussian aEFM \\ m=3, $\sigma=1/2$\end{tabular}} &
  {\color[HTML]{3D3D3D} 83.06±0.12} &
  {\color[HTML]{3D3D3D} 68.99±0.20} &
  {\color[HTML]{3D3D3D} 76.32±0.04} &
  {\color[HTML]{3D3D3D} 63.23±0.06} &
  {\color[HTML]{3D3D3D} 50.83±0.12} \\ \hline
{\color[HTML]{3D3D3D} DeepLabV3+} &
  {\color[HTML]{3D3D3D} 88.24} &
  {\color[HTML]{3D3D3D} 88.79} &
  {\color[HTML]{3D3D3D} 88.52} &
  {\color[HTML]{3D3D3D} 81.14} &
  {\color[HTML]{3D3D3D} 75.10} \\ \hline
\end{tabular}
\end{table}

We also report in Table \ref{tab:linearSVMEnsemble} the best classification results achieved by the ensemble of the SVM classifiers. As it was the case with the ensemble of logistic classifiers, classification performance in the input RGB color space has significantly improved compared to a single classifier performance. Performances of the ensemble of classifiers in space induced by polynomial EFM and Gaussian aEFM are better than performance of the ensemble of classifiers in the input space. However, as with a logistic classifier, it is important to notice that the best performance in terms of micro BACC is achieved by a single classifier trained Gaussian aEFM induced space of order $m=2$. Again, that is important because training an ensemble of classifiers is computationally very demanding.

\begin{table}
\centering
\caption{Performance of the ensemble of 58 SVM classifiers on intraoperative pixel wise classification of adenocarcinoma of a colon in a liver for data in the input RGB color space and space induced by aEFMs. Each classifier is learned on one image from the training set. Results are reported in terms of mean and standard deviation estimated from 10 runs. The testing results shown in boldface are statistically significantly better (with a significance level of 0.05) than the result in the input RGB color space. }
\label{tab:linearSVMEnsemble}
\begin{tabular}{|c|c|c|c|c|c|}
\hline
{\color[HTML]{3D3D3D} \textbf{Model}} &
  {\color[HTML]{3D3D3D} \textbf{SE {[}\%{]}}} &
  {\color[HTML]{3D3D3D} \textbf{SP{[}\%{]}}} &
  {\color[HTML]{3D3D3D} \textbf{BACC {[}\%{]}}} &
  {\color[HTML]{3D3D3D} \textbf{F1 {[}\%{]}}} &
  {\color[HTML]{3D3D3D} \textbf{PPV {[}\%{]}}} \\ \hline
{\color[HTML]{3D3D3D} input RGB color space} &
  {\color[HTML]{3D3D3D} 72.33±0.31} &
  {\color[HTML]{3D3D3D} 78.82±0.22} &
  {\color[HTML]{3D3D3D} 75.58±0.09} &
  {\color[HTML]{3D3D3D} 63.56±0.11} &
  {\color[HTML]{3D3D3D} 56.69±0.18} \\ \hline
{\color[HTML]{3D3D3D} \begin{tabular}[c]{@{}c@{}}polynomial EFM \\ $m=3$, $b \in \{1, \dots , 7\}$ 
\end{tabular}} &
  {\color[HTML]{3D3D3D} \textbf{81.35±0.07}} &
  {\color[HTML]{3D3D3D} 72.72±0.08} &
  {\color[HTML]{3D3D3D} 77.03±0.02} &
  {\color[HTML]{3D3D3D} 64.12±0.03} &
  {\color[HTML]{3D3D3D} 53.33±0.06} \\ \hline
{\color[HTML]{3D3D3D} \begin{tabular}[c]{@{}c@{}}Gaussian aEFM \\ $m=2$, $\sigma \in\{4, 2\sqrt{2}, 2, \sqrt{2}, 1, 1/\sqrt{2}, 1/2, \sqrt{2}/4\}$\end{tabular}} &
  {\color[HTML]{3D3D3D} 74.37±0.23} &
  {\color[HTML]{3D3D3D} \textbf{79.73±0.19}} &
  {\color[HTML]{3D3D3D} \textbf{77.05±0.04}} &
  {\color[HTML]{3D3D3D} \textbf{65.45±0.05}} &
  {\color[HTML]{3D3D3D} \textbf{58.44±0.16}} \\ \hline
\end{tabular}
\end{table}

\section{Discussion}
\label{sec:disc}

Thanks to the advance in computing resources, architectures and expert labeled datasets, deep NN algorithms in medicine can achieve highly accurate predictions (diagnosis). That, however, is associated with the lack of interpretability of features upon which deep NN makes decision \citep{rudin2019stop,tjoa2021asurvey}. In case of high-stakes decisions, it is necessary to understand medical interpretation of predictions made by the algorithm. That is why creation of the interpretable models is advocated in the first place instead of explaining the black box models \citep{rudin2019stop}. Furthermore, in some medical imaging domains, such as intraoperative pixel wise diagnosis of cancer, there is still a lack of experts’ based annotated data necessary for training highly complex deep models. Thus, such models can be prone to overfitting. On the other hand, if nonlinear classification and/or clustering problem is formulated in terms of the less complex kernel-based algorithms, the necessary incremental learning is hard to implement. Thus, clustering and classification of large-scale medical imaging datasets is still a challenging problem.

As opposed to nonlinear classification models, linear models are easily interpretable and explainable when using Shapely additive values to explain individual feature contribution towards model prediction. They are also easily implementable in the incremental learning mode, which is important for large-scale classification problems. Their limitation is a modest diagnostic performance, especially in comparison with deep NN models. That is a consequence of the small model complexity.

\subsection{Linear classification in explicit feature maps induced space}
The central contribution of this paper is the formulation of computationally efficient and interpretable nonlinear classification models for low-dimensional datasets. They are obtained by applying existing linear algorithms in the low-dimensional subspace of the Hilbert space induced by explicit feature maps of small order. Complexity of obtained nonlinear algorithms is still small and models are less prone to overfitting. They remain interpretable in mapping induced Hilbert space. Interpretability is a consequence of the analytical character of the explicit feature maps used for data mapping. Thus, each feature in induced space is a known analytical function of the features from low-dimensional input space. Since linear algorithms are applied to mapped data, they are immediately suitable for incremental learning, and that is of key importance for large-scale classification problems.

\subsection{Clustering and classification of {H\&E} stained images of frozen sections of adenocarcinoma in a liver}
We applied proposed concept to intraoperative pixel wise diagnosis of adenocarcinoma of a colon in a liver from color images of $H\&E$ stained frozen sections. Besides training a single classifier from the training set, we also demonstrated possibility to train ensemble of classifiers by dividing the training set into subsets. Compared to a single linear classifier, the ensembles boost performance of the in the input RGB color space, but also further improve performances in low-dimensional subspace of the Hilbert space induced by EFMs. It is of practical importance that single classifier learned in a space induced by aEFM associated with the Gaussian kernel yielded micro BACC statistically better than the one achieved by the ensemble of linear classifiers in the input space and EFM induced space. That is the case for both SVM classifier and logistic classifier. By using Shapely additive explanation values \citep{lundberg2017unified,aas2021explaining}, we could provide interpretable explanations for features in ten-dimensional space induced by the EFM associated with the polynomial kernel of order $m=2$. We found out that red-green and green-blue features, which do not exist in the input space, yield important contributions towards model predictions. That is interpreted by the role of red-green and green-blue mixtures in formation of blue-purple, pink and white colors that respectively dye nucleus, cytoplasm and glandular structures.

\subsection{Limitations}
Proposed use of linear classifiers in low-dimensional subspace of the Hilbert space induced by aEFMs was motivated by a lack of interpretability of deep NN models and by reduced generalization ability of such when there is a lack of annotated data available for training. That is the case with pixel wise annotated images of $H\&E$ stained frozen sections collected intraoperatively. Even though proposed approach brought a statistically significant performance improvement, in terms of three metrics, when compared with the corresponding models in the input data space, it is still inferior in comparison with deep NN models such as DeepLabV3+. Thus, our approach overcame the existing gap in performance only partially. Arguably, that could be the tradeoff between performance and interpretability.

\section{Conclusion}
In this paper, we presented a new mathematical framework for interpretable nonlinear clustering and classification of large-scale low dimensional medical imaging data. The framework is based on the usage of existing linear models in low-dimensional subspace of the Hilbert space induced by approximate explicit feature maps of low order. Proposed methodology is directly suitable for incremental learning, and that is of key importance for large-scale classification problems. The method was applied to pixel-wise clustering and classification of adenocarcinoma of a colon in a liver from images of $H\&E$ stained specimens of frozen sections. Statistically significant improvements in micro BACC, $\mathrm{F}_1$ score and precision were obtained by linear SVM classifier and linear logistic classifier in mapping induced space in comparison with the performance achieved by the same classifiers in the input data space. The same applies to the results obtained by U-SPEC clustering algorithm. We also discussed interpretability of features in ten-dimensional space induced by EFM associated with a polynomial kernel of order 2. It is conjectured that proposed framework can be applied to clustering and classification of other low dimensional medical datasets such as those collected in PET-CT imaging, MR imaging or multi-phase CT imaging.

\section*{Appendix A. Derivation of explitict feature maps for Gaussian and polynomial kernels}
\label{appendixA}
It is known that function $\kappa(\bm{x},\bm{y})$, that satisfies conditions of the Mercer's theorem \citep{mercer1909functions}, can be written in terms of expansion based on eigenvalues $\{\lambda_n > 0 \}_{n\in\mathbb{N}}$ and eigenfucntions $\{\phi_n(\bm{x})\}_{n\in\mathbb{N}}$ of the compact self-adjoint integral operator on $L^2(\Omega):(L_Kf)(\bm{x})=\int_{\Omega}K(\bm{x},\bm{y})f(\bm{y})dy$:

\begin{equation}
\label{eq:appx1}
    \kappa(\bm{x},\bm{y})=\sum_{n\in\mathbb{N}}\lambda_n\phi_n(\bm{x})\phi_n(\bm{y}).
\end{equation}

\noindent Thanks to the Moore-Aronszjan theorem \citep{aronszajn1950theory}, the function $\kappa(\bm{x},\bm{y})$ can also be written as follows:
\begin{equation}
\label{eq:appx2}
    \kappa(\bm{x},\bm{y})=\left< \phi(\bm{x}),\phi(\bm{y})\right>_{H_k}.
\end{equation}

\noindent Eq. (\ref{eq:appx2}) is known as \textit{kernel trick} \citep{aizerman1964theoretical}. The nonlinear mapping $\phi(\bm{x})$ is called explicit feature map (EFM) and $H_{\kappa}$ denotes a Hilbert space of functions induced by $\kappa(\circ, \bm{x})$. Kernel expansion (\ref{eq:appx1}) implied by Mercer's theorem can be interpreted on the following way \citep{kennedy2013hilbert}:

\begin{equation}
\label{eq:appx3}
\begin{aligned}
\kappa(\bm{x}, \bm{y}) &=\sum_{n \in \mathbb{N}} \sqrt{\lambda_{n}} \phi_{n}(\bm{x}) \sqrt{\lambda_{n}} \phi_{n}(\bm{y}) \\
&=\left\langle\left\{\sqrt{\lambda_{n}} \phi_{n}(\bm{x})\right\}_{n \in \mathbb{N}},\left\{\sqrt{\lambda_{n}} \phi_{n}(\bm{y})\right\}_{n \in \mathbb{N}}\right\rangle_{\ell^{2}} \\
&=\langle\phi(\bm{x}), \phi(\bm{y})\rangle_{\ell^{2}}
\end{aligned}
\end{equation}

\noindent where $\ell^{2}$ denotes feature space. Thus, when $\kappa(\bm{x}, \bm{y})$ is Mercer's kernel EFM $\phi(\bm{x})$ is composed of eigenvalues $\left\{\lambda_{n}\right\}_{n \in \mathbb{N}}$ and eigenfunctions $\left\{\phi_{n}(\bm{x})\right\}_{n \in \mathbb{N}}$ of $\kappa(\bm{x}, \bm{y})$, i.e. $\phi(\bm{x})=\left\{\sqrt{\lambda_{n}} \phi_{n}(\bm{x})\right\}_{n \in \mathbb{N}}$. In kernel-based nonlinear algorithms, one does not need to design particular eigenvalues $\left\{\lambda_{n}\right\}_{n \in \mathbb{N}}$ and eigenfunctions $\left\{\phi_{n}(\bm{x})\right\}_{n \in \mathbb{N}}$, but chooses simple kernel function $\kappa(\bm{x}, \bm{y})$ that satisfies conditions of Mercer's theorem. One such example is the Gaussian kernel: 

\begin{equation}
\label{eq:appx4}
\kappa(\bm{x}, \bm{y})=\exp \left(-\frac{\|\bm{x}-\bm{y}\|_{2}^{2}}{2 \sigma^{2}}\right)
\end{equation}

\noindent To derive analytical expression for $\phi(\bm{x})$ we write (\ref{eq:appx4}) as:

\begin{equation}
\label{eq:appx5}
\kappa(\bm{x}, \bm{y})=\exp \left(-\frac{\|\bm{x}\|_{2}^{2}}{2 \sigma^{2}}\right) \exp \left(-\frac{\|\bm{y}\|_{2}^{2}}{2 \sigma^{2}}\right) \exp \left(-\frac{\langle\bm{x}, \bm{y}\rangle}{\sigma^{2}}\right)
\end{equation}

\noindent By using the multinomial theorem we write:

\begin{equation}
\label{eq:appx6}
\exp \left(-\frac{\langle\bm{x}, \bm{y}\rangle}{\sigma^{2}}\right)=\sum_{k=0}^{\infty} \frac{1}{\sigma^{2 k}} \sum_{|{\boldsymbol{\alpha}} |=k}\left(\begin{array}{l}
k \\
\boldsymbol{\alpha}
\end{array}\right) \bm{x}^{\boldsymbol{\alpha}}  \bm{y}^{\boldsymbol{\alpha}} 
\end{equation}

\noindent In (\ref{eq:appx6}) we used the multi-index notation, i.e. for $\alpha \in \mathbb{N}_{0}^{d}$ we have:

\begin{equation}
\label{eq:appx7}
|\boldsymbol{\alpha}|=\alpha_{1}+\alpha_{2}+\ldots+\alpha_{d}, \boldsymbol{\alpha} !=\alpha_{1} ! \cdot \alpha_{2} ! \cdot \ldots \cdot \alpha_{d} !, \text { and } \bm{x}^{\boldsymbol{\alpha}} =x_{1}^{\alpha_{1}} \cdot x_{2}^{\alpha_{2}} \cdot \ldots \cdot x_{d}^{\alpha_{d}}
\end{equation}

\noindent Also, the binomial coefficient in multi-index notation is defined as:

\begin{equation}
\label{eq:appx8}
\left(\begin{array}{l}
k \\
\boldsymbol{\alpha}
\end{array}\right)=\frac{k !}{(k-|\boldsymbol{\alpha}|) ! \times \boldsymbol{\alpha} !}
\end{equation}

\noindent In direct comparison between the \textit{kernel trick} (\ref{eq:appx2}) and (\ref{eq:appx5})/(\ref{eq:appx6}) we obtain EFM:

\begin{equation}
\label{eq:appx9}
\phi(\bm{x})=\exp \left(-\frac{\|\bm{x}\|_{2}^{2}}{2 \sigma^{2}}\right) \sum_{k=0}^{\infty} \frac{1}{\sigma^{k}} \sum_{|\alpha|=k} \frac{\bm{x}^{\alpha}}{\sqrt{\boldsymbol{\alpha} !}}
\end{equation}

\noindent Occasionally, instead of Gaussian kernel (\ref{eq:appx4}) its "simplification" is used: 

\begin{equation}
\label{eq:appx10}
\kappa(\bm{x}, \bm{y})=\exp \left(-\frac{\|\bm{x}-\bm{y}\|_{2}^{2}}{\sigma^{2}}\right)
\end{equation}

\noindent It is straightforward to derive the analytical expression for EFM:

\begin{equation}
\label{eq:appx11}
\phi(\bm{x})=\exp \left(-\frac{\|\bm{x}\|_{2}^{2}}{\sigma^{2}}\right) \sum_{k=0}^{\infty} \frac{1}{\sigma^{k}} \sum_{|a| = k} \sqrt{\frac{2^{k}}{\boldsymbol{\alpha} !}} \bm{x}^{\alpha}
\end{equation}

\noindent The approximate EFM (aEFM) of order $m, \phi_{m}(\bm{x})$, is obtained when in (\ref{eq:appx9}), respectively (\ref{eq:appx11}), the first summation is carried out from 0 to $m$ :

\begin{equation}
\label{eq:appx12}
\phi_{m}(\bm{x})=\exp \left(-\frac{\|\bm{x}\|_{2}^{2}}{2 \sigma^{2}}\right) \sum_{k=0}^{m} \frac{1}{\sigma^{k}} \sum_{|\alpha|=k} \frac{\bm{x}^{\boldsymbol{\alpha}}}{\sqrt{\boldsymbol{\alpha} !}}
\end{equation}

\noindent We now derive analytical expression for EFM associated with the polynomial kernel:

\begin{equation}
\label{eq:appx13}
\kappa(\bm{x}, \bm{y})=(\langle\bm{x}, \bm{y}\rangle+b)^{m}
\end{equation}

\noindent By using the kernel trick identity, it is to straightforward to derive EFM $\phi(\bm{x})$ for $b=0$ :

\begin{equation}
\label{eq:appx14}
\phi(\bm{x})=\left\{\sqrt{\left(\begin{array}{l}
m \\
\boldsymbol{\alpha}
\end{array}\right)} \bm{x}^{\boldsymbol{\alpha}}\right\}_{|{\boldsymbol{\alpha}}|=m}
\end{equation}

\noindent and for $b>0$ :

\begin{equation}
\label{eq:appx15}
\phi(\bm{x})=\left\{\sqrt{\left(\begin{array}{l}
m \\
\boldsymbol{\alpha}
\end{array}\right) b^{d-|{\boldsymbol{\alpha}}|} \bm{x}^{{\boldsymbol{\alpha}}}}\right\}_{|{\boldsymbol{\alpha}}| \leq m}
\end{equation}

\section*{Declaration of Competeing Interests}
\noindent The authors declare no competing interests or personal relationships that could influence results in reported work. 

\section*{Credit authorship contribution statement}
\noindent \textbf{Dario Sitnik}: Methodology, Software, Visualization, Validation, Formal analysis, Data curation, Writing - Review $\&$ Editing. \\ \textbf{Ivica Kopriva}: Conceptualization, Methodology, Software, Validation, Formal analysis, Writing - Original Draft, Supervision.

\section*{Acknowledgments}
\noindent This work has been supported in part by the Croatian Science Foundation Grant IP-2016-06-5235 and in part by the European Regional Development Fund under the grant KK.01.1.1.01.0009 (DATACROSS).

\bibliography{main}

\end{document}